\documentclass[preprintnumbers,3p,12pt]{elsarticle}

\usepackage{amsmath,amssymb}
\usepackage{mathtools}
\usepackage{bm}
\usepackage{graphicx}
\usepackage{MnSymbol}
\usepackage{url}
\usepackage{hyperref}

\newcommand \bra[1]{\left< {#1} \,\right\vert}
\newcommand \ket[1]{\left\vert\, {#1} \, \right>}

\newcommand{\bea}{\begin{eqnarray}}
\newcommand{\eea}{\end{eqnarray}}
\newcommand{\simgt}{\hbox{ \raise3pt\hbox to 0pt{$>$}\raise-3pt\hbox{$\sim$} }}
\newcommand{\simlt}{\hbox{ \raise3pt\hbox to 0pt{$<$}\raise-3pt\hbox{$\sim$} }}

\newcommand{\alfs}{\alpha_s}

\newcommand{\be}{\begin{equation}}
\newcommand{\ee}{\end{equation}}
\newcommand{\LQ}{\Lambda_{\rm QCD}}
\newcommand{\LMS}{\Lambda_{\overline{\rm MS}}}

\journal{Physics Letters B}

\begin{document}

\begin{flushright}
    \normalsize TU--1114, KEK--TH--2288
\end{flushright}
%\preprint{KYUSHU-HET-210}

\begin{frontmatter}

\title{New method for renormalon subtraction using Fourier transform}

\author{Y.~Hayashi$^a$, Y.~Sumino$^a$ and H.~Takaura$^b$
\vspace*{3mm}
}

\address{
$^a$Department of Physics, Tohoku University,
Sendai, 980--8578 Japan
\\
$^b$Theory Center, KEK, Tsukuba, Ibaraki 305-0801, Japan
}%

\begin{abstract}
To
improve accuracy in calculating QCD effects,
we propose a method for renormalon subtraction in
the context of the operator-product expansion. 
%For a general single-scale observable,
The method enables
%utilizes properties of Fourier transform and 
subtracting renormalons 
of various powers in $\LQ$ 
efficiently and simultaneously
from single-scale observables.
%The method uses Fourier transform for
%suppressing renormalons in the integrand of a one-parameter
%integral expression.
We apply it to different observables and
examine consistency with
theoretical expectations.
\end{abstract}

\end{frontmatter}

In view of an outstanding success of the Standard Model for
particle physics today, it is becoming highly important to
calculate observables of particle physics
with high precision.
This would be needed, among others, to probe the next energy scale 
that dictates the laws of elementary particle physics.
In particular, it is still challenging to calculate various QCD effects accurately,
even though many technologies for
such calculation have been developed over the
past decades.

Use of the operator-product expansion (OPE) in combination
with renormalon subtraction is a way to achieve
high precision calculation of QCD effects.
In the OPE of an observable, ultraviolet (UV)
and infrared (IR) contributions are factorized  into the Wilson coefficients
and nonperturbative matrix elements, respectively.
The former are calculated in perturbative QCD, while the
latter are determined by nonperturbative methods.
It has been recognized, however, that (IR) renormalons
are contained in the respective parts.
IR renormalons are the source of rapid growth of perturbative coefficients,
which lead to bad convergence of perturbative series.
They limit achievable accuracies and induce inevitable uncertainties even when using the OPE framework.
Since a physical observable as a whole should not contain renormalon uncertainties, 
the necessary task is to separate the renormalons
from the respective parts and cancel them.

Historically cancellation of renormalons made a strong impact 
in heavy quarkonium physics.
The perturbative series for the quarkonium energy levels turned out to be
poorly convergent when the quark pole mass was used to express
the levels, following the conventional wisdom.
When they were reexpressed by a short-distance quark mass,
the convergence of the perturbative series improved dramatically.
This was understood as due to cancellation of the ${\cal O}(\LQ)$
renormalons between the pole mass and binding energy \cite{Pineda:id, Hoang:1998nz, Beneke:1998rk}.
A similar cancellation was observed in the 
$B$ meson partial
widths in the semileptonic decay modes \cite{Ball:1995wa,Bigi:1994em,Neubert:1994wq}.
These features were applied in accurate determinations of fundamental
physical constants such as 
the heavy quark masses \cite{Kiyo:2015ufa,Bazavov:2018omf}, some of the 
Cabbibo-Kobayashi-Maskawa matrix elements \cite{Hoang:1998hm,Alberti:2014yda},
and the strong coupling constant $\alfs$ \cite{Bazavov:2012ka}.

Analyses including cancellation of renormalons beyond the ${\cal O}(\LQ)$
renormalon
of the pole mass have started only recently.
The ${\cal O}(\LQ)$ and ${\cal O}(\LQ^3)$ renormalons
of the static QCD potential $V_{\rm QCD}(r)$ were subtracted
%By comparing the renormalon-subtracted leading Wilson coefficient
%with the lattice result,
 and combined with
the corresponding nonperturbative matrix elements 
%\cite{Sumino:2005cq, Takaura:2018lpw, Takaura:2018vcy},
\cite{Takaura:2018lpw},
which were determined by comparing the renormalon-subtracted 
leading Wilson coefficient
with a lattice result.
By the renormalon subtraction,
the perturbative uncertainty of the Wilson coefficient
reduced considerably and the matching range with the lattice
data became significantly wider. 
%Properties of the renormalon-subtracted OPE were
%consistent with theoretical expectations.
%The strong coupling constant
By the matching
$\alfs$ was determined with
good accuracy, which agreed with other measurements.
Also the ${\cal O}(\LQ^4)$ renormalon contained in the lattice
plaquette action was subtracted and absorbed into
the local gluon condensate \cite{Ayala:2020pxq}.

In this paper we generalize the method which was previously established
only for $V_{\rm QCD}(r)$ \cite{Sumino:2005cq}.
In a general case it was difficult to find a simple integral representation of 
renormalons which enables separation of them from the rest.
A transparent way was found to understand the corresponding
mechanism for $V_{\rm QCD}(r)$ using properties of the Fourier transform
\cite{Sumino:2020mxk}.
We are now able to apply it to
a general single-scale observable,
and we present the necessary formulation.
The formulation is simple and easy to compute, once
the standard perturbative series is given for the 
observable.

We 
apply our method to subtract
the following renormalon(s)
in our test analyses:
the ${\cal O}(\LQ^4)$ renormalon of the Adler function;
the ${\cal O}(\LQ^2)$ renormalon of the $B \to X_u \ell \bar{\nu}$ decay width;
the ${\cal O}(\LQ)$ and ${\cal O}(\LQ^2)$ renormalons simultaneously
of the $B$ or $D$ meson mass.
The renormalons of ${\cal O}(\LQ^2)$ in the $B$ decay and 
$B$, $D$ meson masses are subtracted for the first time in this paper.
We show that our results meet theoretical expectations, e.g., good convergence and
consistent behavior with the OPE.
\medbreak

Consider a general dimensionless observable $X(Q)$ 
with a characteristic hard scale $Q$, whose
perturbative expansion is given by $X=\sum_n c_n  \alfs^{n+1}$,
where $\alfs=\alfs(\mu)$ and $\mu$ is the renormalization scale.

An ambiguity induced by renormalons is defined
from the discontinuities of the corresponding singularities in the Borel plane:
%One way to
%define an ambiguity induced by renormalons is
%to use an integral surrounding the discontinuities of the
%corresponding singularities in the Borel plane:
\be
\delta X = \frac{1}{b_0}\int_{C_+ - C_-} \!\!\!\!\!\!\!\!\!\!\!
du \, e^{-u/(b_0\alfs)}  B_X(u) \,,
~~~
 B_X=\sum_n \frac{c_n}{n!}\Bigl(\frac{u}{b_0}\Bigr)^n \,.
\label{ren-by-BI}
\ee
In the complex Borel ($u$) plane
the integral contours $C_\pm(u)$ connect the origin and $\infty\pm i\epsilon$
infinitesimally above/below the positive real axis on which the discontinuities are located.
$b_0=(11-2n_f/3)/(4\pi)$ denotes the one-loop coefficient
of the QCD beta function, where $n_f$ is the number of quark flavors.

The location of a renormalon singularity $u_*$ and the form of 
$\delta X$ due to this singularity
(apart from its overall normalization $N_{u_*}$) can be
determined from the OPE and renormalization-group (RG) equation \cite{Beneke:1998ui}.
Typically  $\delta X(Q)\approx N_{u_*} (\LMS/Q)^{2u_*}$, 
up to systematically calculable
corrections [expressed by 
the anomalous dimension and series expansion in $\alfs(Q)$],
where $\LMS$ denotes the dynamical scale defined in the $\overline{{\rm MS}}$ scheme.
%It is essential for precise QCD predictions to remove
%renormalon uncertainties from theoretical calculations.

To regulate the divergent behavior of $X(Q)$, 
we adopt the usual ``principal value (PV) prescription" 
and take the PV of the Borel resummation integral, that is,
take the  average over the contours $C_\pm(u)$, 
\be
[X(Q)]_{\rm PV}= \frac{1}{b_0}\int_{0,\rm PV}^{\infty}
du \, e^{-u/(b_0\alfs)}  B_X(u)\,.
\label{PV-Bl}
\ee
In this regularization, 
the renormalons are minimally subtracted from $X(Q)$ by definition. 
How to evaluate the PV integral in eq.~(\ref{PV-Bl}) is non-trivial 
in practice with the limited number of known perturbative coefficients.
In this paper we propose a new method to obtain
the PV integral in a systematic approximation
using a different integral representation.

To evaluate $[X(Q)]_{\rm PV}$,
we extend the formulation of ref.~\cite{Sumino:2005cq},
which works for $V_{\rm QCD}(r)$.
In the static QCD potential, renormalons are located
at $u_*=1/2$, $3/2$, $\dots$.
It is known \cite{Hoang:1998nz, Beneke:1998rk, Brambilla:1999qa, Sumino:2020mxk} 
that these renormalons are eliminated (or highly suppressed) and 
the perturbative series exhibits good convergence
in the momentum-space potential,
i.e., the Fourier transform of 
the coordinate-space potential $V_{\rm QCD}(r)$.
Equivalently, $V_{\rm QCD}(r)$ is given by the inverse Fourier integral
of the momentum-space potential, which is (largely) free from the renormalons.
This indicates that the renormalons of $V_{\rm QCD}(r)$
arise from (IR region of) the inverse Fourier transform.
Using the formulation of ref.~\cite{Sumino:2005cq}, 
one can avoid the renormalon uncertainties reviving from the inverse Fourier transform
and  give a renormalon-subtracted prediction for $V_{\rm QCD}(r)$.
This is realized by  proper deformation of the integration contour of the inverse Fourier transform.
In this method, one minimally subtracts the renormalons using Fourier transform, 
and the renormalon-subtracted prediction is actually equivalent to the PV integral~\eqref{PV-Bl}
 as long as the momentum-space potential is free of renormalons.
We propose a generalized method using an analogous mechanism. 
The key to achieving this goal is to find a proper Fourier transform
such that the renormalons of an original quantity are suppressed.

%For a general observable X(Q), ... define a Fourier transform of
%$r^{2 a u'} X$ (where ...) into "momentum ($\tau$) space" as
For a general observable $X(Q)$, let $r=|\vec{x}|=Q^{-1/a}$ and define a Fourier transform of $r^{2au'}X$
(where $a$ and $u'$ are parameters) into ``momentum ($\tau$) space"  as
\be
\tilde{X}(\tau)=\int d^3\vec{x} \, {e^{i\vec{\tau}\cdot\vec{x}}}\,{r^{2au'}}X(r^{-a})
\, ,
~~~(\tau=|\vec{\tau}|) \, ,
 \label{FTX}
\ee 
whose typical energy scale is now $\tau^a$.
Since the Borel  resummation and Fourier transform mutually
commute, it follows that 
\be
\delta\tilde{X}(\tau)=\int d^3\vec{x} \, {e^{i\vec{\tau}\cdot\vec{x}}}\,{r^{2au'}}
\delta X(r^{-a})
\, ,
 \label{FTdelX}
\ee
where  $\delta\tilde{X}$ is defined from $\tilde{X}$ similarly to eq.~(\ref{ren-by-BI}).
Substituting the leading form 
$ N_{u_*} (\LMS/Q)^{2u_*}$ of $\delta X$ to eq.~(\ref{FTdelX}) we obtain
\be
\delta\tilde{X}\approx -\frac{4\pi N_{u_*}\LMS^{2u_*} }{\tau^{3+2a(u_*+u')}}
\sin(\pi a (u_*+u'))\,
\Gamma(2a(u_*+u')+2)
\, .
%\frac{4\pi N_{u_*}\LMS^{2u_*} }{\tau^{3+2a(u_*+u')}}
%\sin(\pi a (u_*+u')+\pi)\,
%\Gamma(2a(u_*+u')+2)
%\, .
 \label{ren-delX}
\ee
In the case of the static potential $X(Q)=r V_{\rm QCD}(1/r)$, with  the choice
$a=1$, $u'=-1/2$,  
$\tilde{X}(\tau)$ reduces to the standard momentum space potential
and the sine factor cancels the leading renormalon at $u_*=1/2$ 
and simultaneously suppresses the
dominant renormalons at $u_*= 3/2, \dots$ \cite{Sumino:2020mxk}.
(In particular, in the large--$\beta_0$ approximation IR renormalons are totally
absent in the momentum-space potential.)
In the case of a general observable $X$, we can adjust the 
parameters $a$ and $u'$
to cancel or suppress the dominant renormalons of $\tilde{X}$.\footnote{
In addition we can vary the dimension of the Fourier transform
to $d^n\vec{x}$. For simplicity we set $n=3$.
}
The level of suppression depends on the observable, but at least the
first two renormalons closest to the origin can always be suppressed.\footnote{
In principle we can
include corrections to eq.~(\ref{ren-delX}) 
generated by anomalous dimension
and higher order terms of $\alfs(Q)$ in $\delta X$, and we can adjust 
additional parameters
to cancel (or suppress more severely) renormalons including the corrections.
On the other hand, 
if these corrections are absent, the current procedure cancels the corresponding
renormalon exactly.
}

We can reconstruct $X(Q) $ by the inverse Fourier transform.
After integrating out the angular variables, naively we obtain
%\be
%X(Q)=\frac{1}{2\pi^2}\int_0^\infty d\tau\, \frac{\sin (\tau r)}{\tau r} \, \tilde{X}(\tau)
%\,.
%\ee
\be
X(Q)=\frac{r^{-2 a u'-1}}{2 \pi^2} \int_0^{\infty} d \tau \, \tau \sin(\tau r) \tilde{X}(\tau) \, .
\label{InvFT-X}
\ee
The left-hand side has renormalons, while the dominant renormalons are
suppressed in $\tilde{X}$ on the right-hand side.
The dominant renormalons are generated by the $\tau$-integral
of logarithms $\log(\mu^2/\tau^{2 a})^n$ at small $\tau$ in the perturbative series for $\tilde{X}(\tau)$.
When we consider resummation of the logarithms by RG alternatively 
(as we will do in practice),
they stem from the singularity of the running coupling constant $\alfs(\tau)$
in  $\tilde{X}$ located on the positive $\tau$ axis.
$\delta X$ is generated by the integral surrounding the
discontinuity of this singularity.
The expected power dependence on $\LMS$ is obtained once we expand
$\sin(\tau r)$ in $\tau$.

We propose to compute the renormalon-subtracted
$X(Q)$ in the PV prescription, $[X(Q)]_{\rm PV}$, in the following way.
We take the principal value of the above integral, that is, take
the average over the contours $C_\pm(\tau)$:
\be
[X(Q)]_{\rm FTRS}=
\frac{r^{-2 a u'-1}}{2 \pi^2} \int_{0,\rm PV}^{\infty} d \tau \, \tau \sin(\tau r) \tilde{X}(\tau) \, .
\label{XFTRS}
\ee
Here, $\tilde{X}(\tau)$ is evaluated by RG-improvement up to
a certain order (see below) and
has a singularity (Landau singularity) on the positive $\tau$-axis.
In the case of $V_{\rm QCD}(r)$,
this quantity coincides with the renormalon-subtracted leading
Wilson coefficient of $V_{\rm QCD}(r)$ used in the analyses \cite{Sumino:2005cq, Takaura:2018lpw}.
%For the explicit way to calculate the principal value integral, see ref.~\cite{Hayashi:future}.  
Since renormalons of $\tilde{X}(\tau)$ are suppressed, the only source of
renormalons in eq.~(\ref{InvFT-X}) is from the integral of the singularity of $\tilde{X}(\tau)$.
Then the PV prescription in Eq.~\eqref{XFTRS}, which minimally regulates the singularity of the integrand 
(or more specifically, that of the running coupling), corresponds to the minimally renormalon-subtracted quantity~\eqref{PV-Bl}.
%Essentially equivalence of eqs.~(\ref{PV-Bl}) and (\ref{XFTRS}) is shown in the appendix of ref.~\cite{Sumino:2020mxk}, while ample numerical
%evidences to support this relation
%for the static potential are presented in the main body of that paper.
We will give an argument for equivalence of eqs.~(\ref{PV-Bl}) and (\ref{XFTRS}) 
up to the N$^4$LL approximation in \cite{Hayashi:future}.
(See also \cite{Sumino:2020mxk}.)
We note that the equivalence holds only when $\tilde{X}(\tau)$
does not have renormalons.
If renormalons remain in $\tilde{X}(\tau)$,
renormalons cannot be removed from the $Q$-space quantity merely by the PV integral in Eq.~\eqref{XFTRS}, 
which only regulates the Landau pole of the running coupling.
Hence, the renormalon suppression in $\tau$-space quantity [cf. Eq.~\eqref{FTdelX}]
is crucial for renormalon subtraction.

$\tilde{X}(\tau)$ in the N$^k$LL approximation is calculated in the
following manner.
From the coefficients of the series up to 
$k$-th order perturbation $X(Q)=\sum_{n=0}^{k}c_n\alfs(Q)^{n+1}$, 
$\tilde{X}$ is given by
\be
\tilde{X}(\tau)\to\tilde{X}^{(k)}(\tau)=\frac{4\pi}{\tau^{3+2au'}}
\sum_{n=0}^k \tilde{c}_n(0)\,\alfs(\tau^a)^{n+1}
\, ,
\label{FTXpert}
\ee
where $\tilde{c}_n(L_\tau)$ is defined by the following relation
\be
F(\hat{H},L_\tau)\sum_{n=0}^\infty c_n\alfs^{n+1}=\sum_{n=0}^\infty \tilde{c}_n(L_\tau)\alfs^{n+1}\,,
\label{relcandtildec}
\ee
\be
F(u,L_\tau)=-\sin\left(\pi a(u+u')\right)\Gamma\left(2a(u+u')+2\right)e^{L_\tau u}\,.
\ee
Here, $L_\tau=\log(\mu^2/\tau^{2a})\,,\,
\hat{H}=-\beta(\alfs)\frac{\partial}{\partial \alfs}\,$.
%The function $F$ is come from the Fourier transform of the factor $(\mu^2 r^{2a})^{\hat{H}}$.
Thus, $\tilde{c}_n$ is given explicitly by the coefficients of the original series 
$c_0\,,\,c_1\,,\,\cdots\,,\, c_n$ as
\bea
&&
\tilde{c}_0(L_\tau)=F(0,0)c_0\,,\,
\\ &&
\tilde{c}_1(L_\tau)=F(0,0)c_1+\partial_u F(0,L_\tau)b_0c_0\,,\,
\\ &&
~~~~~\vdots~~\,.
\nonumber
\eea
The relation (\ref{relcandtildec}) is a straightforward consequence of
the Fourier transform.
Since the renormalons in $\tilde{X}(\tau)$ are suppressed
and its perturbative series has a good convergence, 
it is natural to perform RG improvement in the $\tau$ space, and
a higher order (large $k$) $\tilde{X}^{(k)}(\tau)$ would be
a more accurate approximation of $\tilde{X}(\tau)$.
Accuracy tests by going to higher orders will be given in the
test analyses below, 
where we estimate a higher order effect by using the 5-loop coefficient of 
the QCD beta function.

\begin{figure}
\begin{center}
\includegraphics[width=5cm]{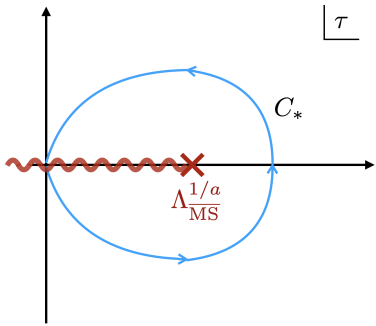}
\end{center}
\caption{Integration contour $C_*$ in Eq.~\eqref{Xpow}.
The cut singularity, shown by the wavy line, is caused by the Landau pole
of the running coupling.}
\label{fig:contour}
\end{figure}

In numerical evaluation of eq.~(\ref{XFTRS}) it is useful to
decompose 
the PV integral into two parts $X_0(Q)$ and $X_{\rm pow}(Q)$
by deforming the integral contour in the complex plane:
\be
[X(Q)]_{\rm FTRS}=\bar{X}_0(Q)+\bar{X}_{\rm pow}(Q)\,,\,
\label{XFTRS-decomp}
\ee
\be
\bar{X}_0(Q)=\frac{-r^{-2 a u'-1}}{2\pi^2} \int_0^{\infty} dt \, t\, e^{-tr}\,{\rm Im}\,[\tilde{X}(\tau=it)]  \,,
\label{X0}
\ee
\be
\bar{X}_{\rm pow}(Q)=\frac{r^{-2 a u'-1}}{4\pi^2 i}\int_{C_*} d \tau \, \tau  \cos(\tau r) \tilde{X}(\tau)\,.
\label{Xpow}
\ee
The integration contour $C_*$ is shown in Fig.~\ref{fig:contour}. 
$\bar{X}_0$ is defined as the integral on the imaginary axis in the upper half $\tau$ plane ($\tau=it$), 
where $e^{-i\tau r}$ turns to a damping factor $e^{-tr}$.
$\bar{X}_{\rm pow}$ can be expanded by $r=Q^{-1/a}$ 
once the expansion of $\cos(\tau r)$ is performed in $\tau$,
and the coefficients of this power series are real.\footnote{
When we take the integration contour as $C_{\pm}$ instead of the PV integral in eq.~\eqref{XFTRS},
we also have power dependence with imaginary coefficients whose sign depends on which contour is chosen.
The power series with imaginary coefficients is identified as renormalon uncertainties.
They do not appear in eq.~\eqref{XFTRS}, where the average over $C_{\pm}$ is taken.
\label{footnote:3}
}
It should be expanded at least to the order of the eliminated renormalon.
Then eq.~(\ref{XFTRS-decomp}) gives the renormalon-subtracted prediction 
of a general observable $X(Q)$
with an appropriate power accuracy of $1/Q$.\footnote{
The results scarcely change by varying the truncation order
beyond the minimum necessary order, 
in the tested range of $Q$ in the examples below.
}
[In the case of the static potential,
eq.~\eqref{X0} is equal to $U_1(r)$ of eq.~(61) in \cite{Sumino:2005cq},
while eq.~\eqref{Xpow} is equal to the renormalon-free part
${\cal A}/r + {\cal C}r + \cdots$ of eq.~(63) in the same paper
with $Q=1/r$.]

The advantages of our method can be stated as follows.
First, our formulation to subtract renormalons works without knowing 
normalization constants $N_{u_*}$ of the renormalons to be subtracted,
following the above calculation procedure.
In other methods \cite{Lee:2002sn,Ayala:2019uaw,Ayala:2019hkn,Ayala:2020odx,Takaura:2020byt},
in order to subtract renormalons 
one needs normalization constants $N_{u_*}$ of the corresponding renormalons.
Normalization constants of renormalons far from the origin are generally difficult to estimate.
Although we certainly need to know large order perturbative series to improve the accuracy of renormalon-subtracted results,
the above feature of our method practically facilitates subtracting multiple renormalons even with 
small number of known perturbative coefficients. 
%In our method, we can achieve renormalon subtraction without explicit results of normalization constants,
%and  is encoded  
%In our method we can automatically obtain perturbative series with good convergence 
%by moving to the Fourier transformed ($\tau$-space) quantities, according to eq.~\eqref{ren-delX},
%without the explicit result of normalization constants.
Secondly, we can give predictions free from the unphysical singularity 
around $Q \sim \LMS$ caused by the running of the coupling, 
in the same way as the previous study of $V_{\rm QCD}(r)$ \cite{Sumino:2005cq}.
Since renormalons and the unphysical singularity are the main sources destabilizing
perturbative results at IR regions, the removal of these factors 
is a marked feature of our method.

We will test validity of the above method 
(renormalon subtraction using Fourier transform: 
``FTRS method'') by applying it to different observables in the
following.\footnote{
As seen in 
eq.~(\ref{ren-delX}), the Fourier transform 
generates artificial UV renormalons in $\tilde{X}$.
They are Borel summable, and we perform the Borel summation 
whenever the induced UV renormalons are located closer to the
origin than the IR renormalons of our interest
 \cite{Hayashi:future}.
}
%When we compare to the calculation without renormalon subtraction,
%``fixed-order calculation'' means that we fix the scale $\mu$ at
%the minimal-sensitivity scale of the corresponding observable
%for a given value of $Q$;
%``RG-improved calculation'' means that we fix the scale at $\mu=Q$.
\medbreak
%\subsection*{Adler function}
\noindent
{Adler function}
\medbreak
The Adler function is defined from the photon vacuum polarization function
$\Pi(-Q^2)$ in the Euclidean momentum region ($Q^2>0$):
\bea
&&D(Q^2)=12\pi^2Q^2\frac{d}{dQ^2}\Pi(-Q^2) 
\nonumber\\
&&~~~~~~~~~
=Q^2\int_0^\infty \!\!\! ds\,
\frac{R(s)}{(s+Q^2)^2}\,.
\label{disprel}
\eea
The second line shows that it is also expressed by
the $R$-ratio through dispersion relation.
The OPE is given by
\be
D(Q^2)=C_1 + 2\pi^2\sum_fQ_f^2C_{GG}\,\frac{\bra{0} \frac{\alpha_s}{\pi} G^{a\mu\nu}G^a_{\mu\nu}\ket{0}}{Q^4}
+ \cdots ,
\ee
where $C_1$ and $C_{GG}$ denote the Wilson coefficients
of the operators $\boldmath \bf{1}$ and $G^{a\mu\nu}G^a_{\mu\nu}$,
respectively.
For simplicity, we omit ${\cal O}(\alpha_s)$ correction to $C_{GG}$ ($C_{GG}=1$).
%\be
%D(Q^2)=C_1 + C_{GG}\,\frac{\bra{0} \frac{\alpha_s}{\pi} G^{a\mu\nu}G^a_{\mu\nu}\ket{0}}{Q^4}
%+ \cdots ,
%\ee
%where $C_1$ and $C_{GG}$ denote the Wilson coefficients
%of the operators $\boldmath \bf{1}$ and $G^{a\mu\nu}G^a_{\mu\nu}$,
%respectively.
We set $n_f=2$ and the quark masses to zero,
hence we ignore the quark condensate $m_i\langle \bar{\psi}_i{\psi_i} \rangle$.

We apply FTRS to the leading Wilson coefficient $C_1$ and
subtract the renormalon at $u=2$, i.e., at order $(\LQ/Q)^4$, so that
the local gluon condensate $\langle (\alpha_s/\pi) G^{a\mu\nu}G^a_{\mu\nu}\rangle$
is also well defined without renormalon uncertainty of the same order of
magnitude.
We choose $a=1/2$, $u'=-2$
such that the renormalons at $u=2, 4, 6, \cdots$ are suppressed.
$\tilde{C}_1(\tau)
=\sum_{n=0}^4 \tilde{d}_n \alfs(\tau)^n$ in eq.~(\ref{XFTRS})
is readily obtained from the NNNLO perturbative calculation of
the Adler function \cite{Baikov:2012zn}.
Here, $\alfs(\tau)$ denotes the 5-loop running coupling constant
in the $\overline{\rm MS}$ scheme \cite{Baikov:2017ayn}. (For simplicity we basically use the 5-loop running constant.)

We change the scale by a factor 2 or $1/2$ from $\mu=\tau$ in 
$\tilde{C}_1(\tau)$.
Consistently with the theoretical expectation,
we verify that the scale dependence of $C_1^{\rm FTRS}$
is considerably smaller than the fixed-order 
or RG-improved calculation without renormalon subtraction,
especially at low energy region.

\begin{figure}[t]
\begin{center}
\includegraphics[width=7.5cm]{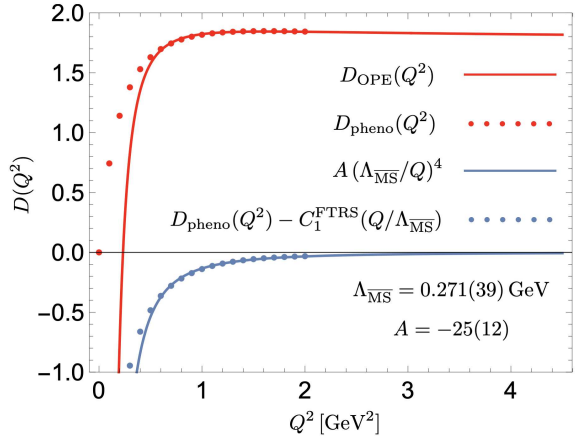}
\end{center}
\vspace*{-5mm}
\caption{
Comparison of Adler function by OPE in FTRS method
and that by phenomenological determination.
They agree reasonably well. 
%$A$ and $\LMS$ in $D_{\rm OPE}$
%are determined by a fit
%to minimize the difference.
}
\label{Fig:comp-AdlerFn}
\vspace*{-4mm}
\end{figure}

As a phenomenological input in the $n_f=2$ case, we
use the phenomenological model \cite{Bernecker:2011gh} for $R(s)$,
after projecting it to only the $n_f=2$ sector.
It is inserted into eq.~(\ref{disprel}) to obtain $D_{\rm pheno}$.
This is compared with 
\be
D_{\rm OPE}(Q^2)=C_1^{\rm FTRS}(Q/\LMS)+A\,\biggl(\frac{\LMS}{Q}\biggr)^4
\,.
\ee
Here, the parameters $A=-25(12)$ and $\LMS=271(39)$~MeV are determined
by a fit to minimize $|D_{\rm pheno}-D_{\rm OPE}|$
in the range $0.6~\text{GeV}^2 \leq Q^2 \leq 2~\text{GeV}^2$;
see Fig.~\ref{Fig:comp-AdlerFn}. (The errors inside the brackets are estimated purely perturbatively by
changing the scale by a factor 2 or $1/2$ from $\mu=\tau$ in 
$\tilde{X}(\tau)$, for a reference.)
In this difference there scarcely remains a room for, e.g., a term
proportional to $(\LMS/Q)^2$.
To improve accuracy by going to higher orders,
it may be important to deal with the $u=-1$ UV renormalon of the Adler function properly.
This speculation is based on our analyses with the 5-loop perturbative coefficient
in the large-$\beta_0$ approximation.

%We estimate that the error of $A$ is dominated by that of
%the perturbative series of $C_1^{\rm FTRS}$, which is seemingly
%converging well.
%This is based on our examination to include the part of the next-order
%correction coming from the already known 5-loop coefficient of the
%QCD beta function \cite{Baikov:2017ayn}.
%The fit results in
%$A=-25.8(6.6)$ and $\LMS=269(17)$~MeV with reduced perturbative
%errors.
%This feature is consistent with the expectation that the $u=2$ renormalon has been
%subtracted.

The above first analysis is fairly crude, 
with unknown uncertainties
included in the model cross section, etc.\footnote{
For instance, the UV renormalons (included in the Adler function on its own) may
give non-negligible contributions, and we have not taken them into account.
}
Nevertheless, it may be informative to compare
the above result with the determination $\LMS^{(2)}=310(20)$~MeV 
in the two-flavor lattice simulation \cite{Aoki:2019cca}, with which we observe a rough consistency.
\medbreak
%\subsection*{\boldmath $B \to X_u \ell \bar{\nu}$ decay width}
\noindent
$B \to X_u \ell \bar{\nu}$ decay width
\medbreak
We consider the decay width of the process $B \to X_u \ell \bar{\nu}$.
The OPE is given within Heavy Quark Effective Theory (HQET)
as an expansion in $1/m_b$ \cite{Manohar:2000dt}:
\be
\Gamma = \Gamma_0 \biggl[
\gamma_1 -\frac{\mu_\pi^2}{2m_b^2} + \frac{\mu_G^2}{2m_b^2} 
+{\cal O}(m_b^{-3})
\biggr] \,,
\ee
where $\Gamma_0$ denotes the partonic decay width
without QCD corrections.
$\gamma_1$ denotes the Wilson coefficient of the
operator $\boldmath \bf{1}$, and
the ${\cal O}(\LQ^2)$ nonperturbative
matrix elements are denoted as
\be
\mu_\pi^2 = \bra{B} h_b^\dagger \vec{D}^2 h_b \ket{B}
\,,~~~
\mu_G^2 = \bra{B} h_b^\dagger \vec{\sigma}\cdot g \vec{B} h_b \ket{B}
\,.
\ee
For simplicity we omit the Wilson coefficients 
multiplying these
matrix elements \cite{Becher:2007tk,Alberti:2013kxa}.

$\Gamma_0$ is proportional to $m_b^5$, and 
it is known that the $u=1/2$ renormalon is canceled
when a short-distance $b$-quark mass 
is used
instead of the pole mass to express the decay width $\Gamma$ \cite{Bigi:1994em}.
Thus, the leading renormalon of $\gamma_1$ is at $u=1$.

We apply FTRS to $\gamma_1$ and
subtract 
the dominant part of 
this renormalon, such that
$\mu_\pi^2$ and $\mu_G^2$ 
are also well defined.
%This is the first trial to subtract the $u=1$ renormalon.
We choose $a=1$, $u'=-1$, hence
renormalons at $u=1, 2, 3, \cdots$ are suppressed.
$\tilde{\gamma}_1(\tau)
=\sum_{n=0}^2 \tilde{s}_n \alfs(\tau)^n$
is derived from the two-loop calculation of
$\Gamma$ \cite{Pak:2008cp}, after expressing it by the $\overline{\rm MS}$
mass $\overline{m}_b$.

\begin{figure}[t]
\begin{center}
\includegraphics[width=7.5cm]{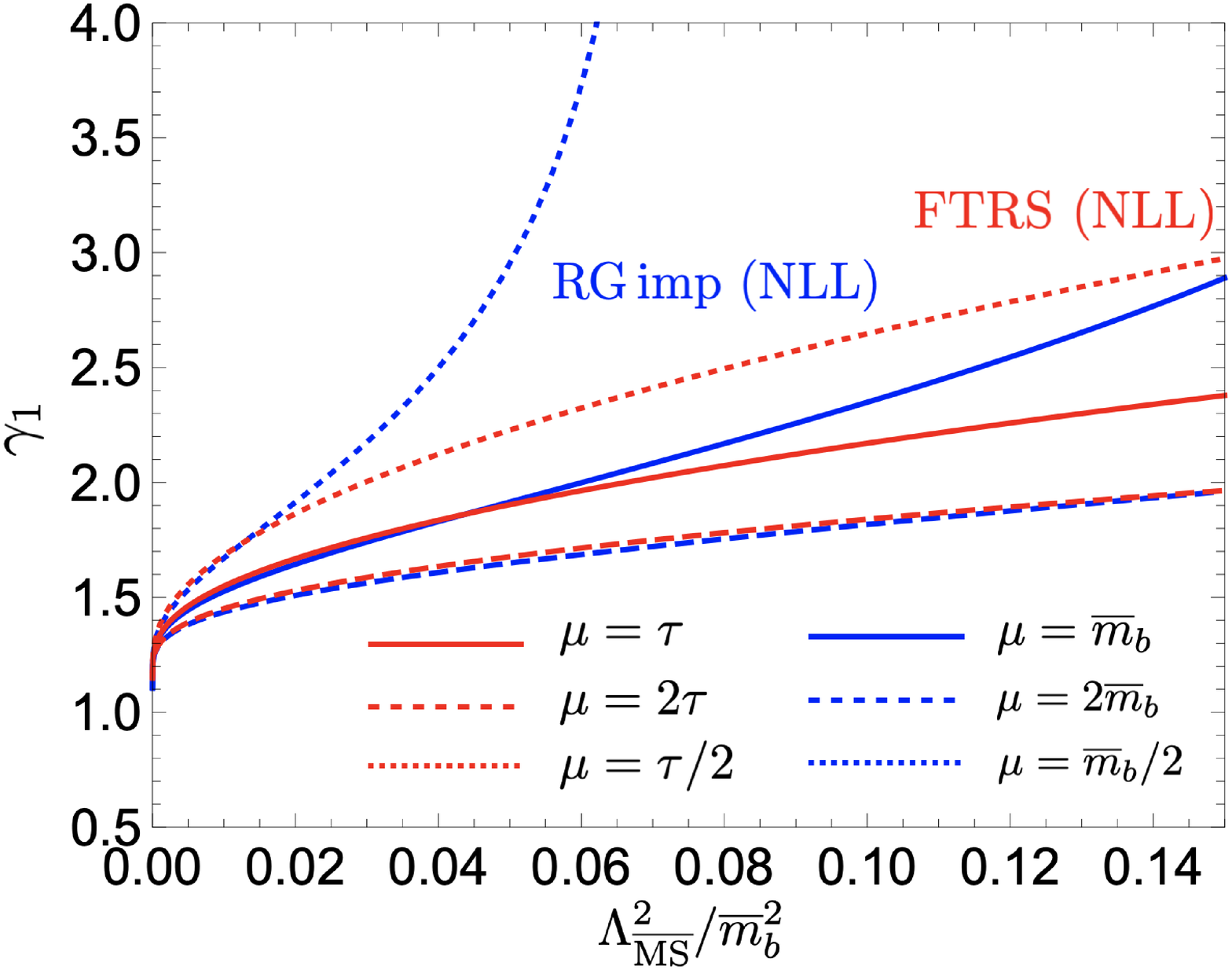}
\end{center}
\vspace*{-5mm}
\caption{
Comparison of
the decay width for $B \to X_u \ell \bar{\nu}$ given by
$\gamma_1^{\rm FTRS}$ and by
RG improvement without renormalon subtraction
(both at NLL and expressed by $\overline{m}_b$).
The former is less scale dependent than the latter
at $\LMS^2/\overline{m}_b^2\simgt 0.02$.
}
\label{Fig:comp-Bdecaywidth}
\vspace*{-4mm}
\end{figure}

We change the scale by a factor 2 or $1/2$ from $\mu=\tau$ in 
$\tilde{\gamma}_1(\tau)$.
For a hypothetically small value $\overline{m}_b \sim 1$--1.5~GeV,
the scale dependence of $\gamma_1^{\rm FTRS}$
is smaller than the fixed-order 
or RG-improved calculation without subtraction of the $u=1$ renormalon;
see Fig.~\ref{Fig:comp-Bdecaywidth}.
However, in the relevant region $\overline{m}_b \approx 4$~GeV,
there are no significant differences in the scale dependence.

The scale dependence of $\gamma_1^{\rm FTRS}$ at the current 
accuracy is still rather
large, where only the corrections up to ${\cal O}(\alfs^2)$ are known.
We expect a healthy convergence behavior of $\gamma_1^{\rm FTRS}$
since we subtracted the
dominant renormalons.
We estimate higher order results by 
including log dependent terms, dictated by RG equation, at the 5-loop level.
We consider RG improvement for perturbative series in $\tau$ space rather than $\bar{m}$ space, 
in accordance with the concept that there are no IR renormalons in the $\tau$-space quantity.
(Note that RG improvements in two spaces are not equivalent when we do not know
exact perturbative coefficients. 
In this treatment, the $u=1$ renormalon is induced in the perturbative series in $\bar{m}$ space.)
Then, a large uncertainty arises in the result of the usual RG improvement.
By FTRS, we obtain a significantly small uncertainty thanks to the renormalon subtraction,
while the analyses are base on the same perturbative series.
However, in order to have an insight into the true size of the $u=1$ renormalon,
we need more terms of the perturbative series.

%We examine it by including those parts proportional to the 
%beta functions, determined by RG, up to 5 loops.
%The scale dependence reduces considerably, indicating convergence.
%The dependence is also somewhat smaller than the
%RG-improved calculation without renormalon subtraction
%(otherwise with the same approximation),
%even at $\overline{m}_b \approx 4$~GeV.
%These features indicate that the contribution of the
%$u=1$ renormalon is not very large for this observable, but
%we need more terms of the perturbative series
%in order to be more conclusive.
\medbreak
%\subsection*{\boldmath $M_B$ and $M_D$}
\noindent
$M_B$ and $M_D$
\medbreak
In HQET the mass of a heavy-light system
$H=B^{(*)}, D^{(*)}$ is given as an expansion in $1/m_h$
($h=b,c$) as \cite{Falk:1992wt}
\be
M_H=m_h + \bar{\Lambda} + \frac{\mu_\pi^2}{2m_h} + w(s)\frac{\mu_G^2}{2m_h}
+{\cal O}(m_h^{-2})
\,.
\ee
Here, $m_h$ denotes the pole mass of $h$;
$\bar{\Lambda}$ represents the contribution 
of ${\cal O}(\LQ)$ from light degrees of freedom;
$s=0,1$ denotes the spin of $H$, and $w(0)=-1$, $w(1)=1/3$.
For simplicity we omit the Wilson coefficients 
multiplying $\mu_\pi^2$ and $\mu_G^2$.
% \cite{Grozin:2007fh}.
By the heavy quark symmetry
$\bar{\Lambda}$, $\mu_\pi^2$ or $\mu_G^2$ are
%defined from light degrees of freedom 
%in the theory with $(n_\ell, n_h)=(3,2)$ with
%expansion around $m_h \to \infty$, and hence they 
common for $h=b,c$.
The renormalons in $m_h$ become
manifest when we express it by a short-distance mass
(we use the $\overline{\rm MS}$ mass $\overline{m}_h$).

We apply FTRS to $m_h$ and
subtract 
the renormalons at $u=1/2$ and $u=1$.
We choose $a=2$, $u'=-1/2$, hence
renormalons at $u=1/2, 1,
%$ are canceled and those at $u=
3/2, 2, \cdots$ are suppressed.
Projecting out $\mu_G^2$, we define
\bea
&&
\overline{M}_{H,{\rm OPE}}
\equiv \frac{1}{4} \bigl( M_{H,{\rm OPE}}^{s=0} + 3 M_{H,{\rm OPE}}^{s=1} \bigr)
\nonumber\\ &&
~~~~~~~~~~~
\equiv m_h^{\rm FTRS} + \bar{\Lambda}^{\rm FTRS} + 
\frac{(\mu_\pi^2)^{\rm FTRS}}{2 m_h^{\rm FTRS}}
\,.
\label{MHbar-OPE}
\eea
Here, $m_h^{\rm FTRS}$ denotes the principal value of 
the pole mass (expressed by 
the $\overline{\rm MS}$ mass, calculated in the FTRS method).

First we examine the scale dependence of $m_h^{\rm FTRS}$
and compare with the fixed-order and RG-improved
calculation without renormalon subtraction.
We use the perturbative series up to ${\cal O}(\alfs^4)$ for 
pole-$\overline{\rm MS}$ mass relation \cite{Marquard:2015qpa,Marquard:2016dcn}.
Since the contribution of the $u=1/2$ renormalon is known to be significant for $m_h$,
(only) in this comparison, $d m_h/d\overline{m}_h$
is used to cancel the $u=1/2$ renormalon.
There have been estimates that the contribution of the $u=1$
renormalon is small \cite{Ayala:2019hkn}. 
(For example, it is absent in the large-$\beta_0$ approximation.)
Consistently with such estimates we observe no significant
difference in the scale dependence if $\overline{m}_h\simgt 1$~GeV, 
even after subtracting the
$u=1$ renormalon.

We determine $\bar{\Lambda}$ and $(\mu_\pi^2)^{\rm FTRS}$
in eq.~(\ref {MHbar-OPE}) by comparing $\overline{M}_{H,{\rm OPE}}$ to the experimental
values of $\overline{M}_B$ and $\overline{M}_D$.
In this analysis we include non-zero charm mass effects to the bottom mass 
(and non-decoupling effects from bottom to the charm mass) 
up to ${\cal O} (\alfs^3)$ \cite{Fael:2020bgs}.
We use the values of $\overline{m}_b$, $\overline{m}_c$, $\alfs$
from the Particle Data Group as input parameters \cite{Zyla:2020zbs}.
We obtain
\bea
&&
\bar{\Lambda}_{\rm FTRS}=0.495(15)_\mu (49)_{\overline{m}_b} (12)_{\overline{m}_c}
(13)_{\alfs} (0)_{\rm f.m.}\text{GeV} \label{Lambdabarfinal}
\,, 
\\&&
(\mu_\pi^2)_{\rm FTRS}=-0.12(13)_\mu (15)_{\overline{m}_b} (11)_{\overline{m}_c}
(4)_{\alfs} (0)_{\rm f.m.}\text{GeV}^2, \label{myupifinal}
\eea
%\bea
%&&
%\bar{\Lambda}^{\rm FTRS}=0.445
%%\nonumber\\&&
%%~~~~~~~
%(20)_\mu (49)_{\overline{m}_b} (12)_{\overline{m}_c}
%(16)_{\alfs} (30)_{\rm f.m.}[\text{GeV}]
%\,,
%\\&&
%(\mu_\pi^2)^{\rm FTRS}=-0.06
%%\nonumber\\&&
%%~~~~~~~
%(18)_\mu (15)_{\overline{m}_b} (11)_{\overline{m}_c}
%(5)_{\alfs} (10)_{\rm f.m.}[\text{GeV}^2]
%\,,
%\eea
where the errors denote, respectively, that from the scale dependence
for $\mu=2\tau$ or $\mu=\tau/2$,
from the errors of the input $\overline{m}_b$, $\overline{m}_c$, $\alfs$,
and from 
the finite $m_c$ corrections (non-decoupling bottom effects) in loops for $m_b$ ($m_c$).
%$\overline{M}_{B,{\rm OPE}}$.

The error from the scale dependence is a measure of perturbative
uncertainty. 
From eq.~\eqref{ren-delX} and our parameter choice, 
the renormalons at $u=1/2$ and $1$ should be removed in our result.
For $\bar{\Lambda}$, about 3 per cent error shows successful subtraction of the $u=1/2$ renormalon. 
On the other hand, the scale dependence of $(\mu_\pi^2)^{\rm FTRS}$
is not smaller than ${\cal O}(\LQ^2)$.
We estimate that this is not because of the contribution from the
$u=1$ renormalon but due to the insufficient number of known terms
of the perturbative series.
%It particularly induces non-small uncertainty to 
%the ${\cal O}(\LMS^2/m^{\rm FTRS}_h)$ term in the renormalon-subtracted prediction.
Examination using the 5-loop beta function
combined with an estimated 5-loop coefficient in the large-$\beta_0$ approximation
indeed indicates that this uncertainty can be reduced at higher order
[$\bar{\Lambda}^{\rm (est)}_{\rm FTRS}=0.488(4)_\mu\,\text{GeV}$
and 
$(\mu_\pi^2)^{\rm (est)}_{\rm FTRS}=-0.09(7)_\mu\text{GeV}^2$].
\medbreak

In summary, in all the above analyses we observed good consistency with
theoretical expectations.
In particular, from the $B,\,D$ meson masses
we obtained
\bea
&&
\bar{\Lambda}_{\rm FTRS}=0.495\pm0.053~\text{GeV} 
\,, 
~~~~~
(\mu_\pi^2)_{\rm FTRS}=-0.12\pm 0.23~\text{GeV}^2
\,,
\label{resLambdamu2final2}
\eea
by subtracting ${\cal O}(\LQ)$ and ${\cal O}(\LQ^2)$ renormalons
simultaneously for the first time.
In general, it is desirable to know more terms of the
relevant perturbative series in order to 
make conclusive statements about the effects of
subtracting renormalons  beyond the ${\cal O}(\LQ)$ renormalon.
The above analyses also show that the 5-loop QCD beta function is
a crucial ingredient in improving accuracies.
We anticipate that the FTRS method can be
a useful theoretical tool for precision QCD calculations 
in the near future.

%We admit that the above first analyses are still premature
%for determining various parameters and
%nonperturbative matrix elements
%precisely.
%It is desirable that we know more terms of the
%relevant perturbative series.
%%Lack of the number of known terms of the relevant 
%%perturbative series is a major reason.
%The analyses also show that the 5-loop QCD beta function is
%a crucial ingredient in improving accuracies.
%Since as far as we could examine we
%observed consistency with theoretical expectations,
%we anticipate that the FTRS method can be
%a useful theoretical tool for precision QCD calculation 
%in the near future.
%

%\medbreak
\section*{Acknowledgements}
Y.H. acknowledges support from GP-PU at Tohoku University. 
The work of Y.H. was also supported in part by Grant-in-Aid for JSPS Fellows (No. 21J10226) from MEXT, Japan. 
The works of Y.S.\ and H.T., respectively, were supported in part by Grant-in-Aid for
scientific research (Nos.\  20K03923 and 19K14711) from
MEXT, Japan.
\vspace{5mm}\\
{\it
Note added:
After completion of this work, we learned that 
the ${\cal O}(\alfs^3)$ corrections to the $B$ meson
semileptonic decay width has recently been computed
in expansion in $(1-m_c/m_b)$ \cite{Fael:2020tow}.
We will present the study including this effect
in \cite{Hayashi:future}.
}

\begingroup\raggedright\endgroup

\end{document}